\documentclass{emulateapj}
\def\Fref#1{Figure~\ref{Fig:#1}}

\newcommand{\altm} {\altaffilmark}
\newcommand{\psr}  {J0737$-$3039}

\newcommand{\hh}   {^{\mathrm h}}  
\newcommand{\mm}   {^{\mathrm m}}  
\newcommand{\dd}   {^\circ}  
\newcommand{\am}   {^\prime} 
\newcommand{\persec}{s$^{-1}$}
 
\shorttitle{Pulsed X-ray Emission from the Double Pulsar}
\shortauthors{Chatterjee et al.}
\slugcomment{Submitted to the Astrophysical Journal}

\begin{document}
\title{Pulsed X-ray Emission from Pulsar A in the Double Pulsar System
  J0737$-$3039}  
\author{S. Chatterjee\altm{1,2}, B. M. Gaensler\altm{1,2}, A. Melatos\altm{3},
 W. F. Brisken\altm{4}, B. W. Stappers\altm{5,6}}

\begin{abstract}
The double pulsar system \psr\ is not only a test bed for General
Relativity and theories of gravity, but also provides a unique
laboratory for probing the relativistic winds of neutron stars. Recent
X-ray observations have revealed a point source at the position of the
\psr\ system, but have failed to detect pulsations or orbital
modulation. Here we report on {\em Chandra X-ray Observatory} High
Resolution Camera observations of the double pulsar.  We detect deeply
modulated, double-peaked X-ray pulses at the period of PSR~\psr A,
similar in appearance to the observed radio pulses. The pulsed
fraction is $\sim 70\%$.  Purely non-thermal emission from pulsar A
plausibly accounts for our observations.  However, the X-ray pulse
morphology of A, in combination with previously reported spectral
properties of the X-ray emission, allows the existence of both
non-thermal magnetospheric emission and a broad sinusoidal thermal
emission component from the neutron star surface.  No pulsations are
detected from pulsar B, and there is no evidence for orbital
modulation or extended nebular structure.  The absence of orbital
modulation is consistent with theoretical expectations of a
Poynting-dominated relativistic wind at the termination shock between
the magnetosphere of B and the wind from A, and with the small
fraction of the energy outflow from A intercepted by the termination
shock.
\end{abstract}


\altaffiltext{1}{School of Physics A29, The University of Sydney, NSW
  2006, Australia; schatterjee, bgaensler@usyd.edu.au}
\altaffiltext{2}{Harvard-Smithsonian Center for Astrophysics, 60
  Garden Street, Cambridge, MA 02138}
\altaffiltext{3}{School of Physics, University of Melbourne,
  Parkville VIC 3010, Australia; amelatos@physics.unimelb.edu.au} 
\altaffiltext{4}{National Radio Astronomy Observatory, P.O. Box O,
   Socorro, NM 87801; wbrisken@aoc.nrao.edu} 
\altaffiltext{5}{Stichting ASTRON, Postbus 2, 7990 AA Dwingeloo, The
  Netherlands; stappers@astron.nl} 
\altaffiltext{6}{Astronomical Institute ``Anton Pannekoek'',
  University of Amsterdam, Kruislaan 403, 1098 SJ Amsterdam, The
  Netherlands}

\section{Background}

Binary neutron star systems are rare, and even among them, the double
pulsar system \psr\ is extraordinary, since both the neutron stars are
detected as radio pulsars.  The system consists of the recycled
22.7~ms pulsar ``A'' \citep{bdp+03} and the young 2.8~s pulsar ``B''
\citep{lbk+04}, in a 2.454~hr eccentric ($e=0.09$) binary orbit which
happens to be nearly edge-on to us.  As well as being a test bed for
General Relativity and theories of gravity \citep[e.g.][]{ksm+06}, the
double pulsar is rich in observational phenomena, including a short
eclipse of A by the magnetosphere of B and orbital modulation of the
radio flux of B due to the influence of A \citep{lbk+04}.  The
individual pulses from B show drifting features due to the impact of
the low-frequency electromagnetic wave in the relativistic wind from A
\citep{mkl+04}, while the eclipse of A is modulated at half the
rotational period of B \citep{mll+04}.  Clearly, the two neutron stars
have both gravitational and electromagnetic interactions with each
other, and the double pulsar system should provide a unique laboratory
to investigate the interactions between the magnetospheres and
relativistic winds of the two pulsars.

In this context, the detection of X-ray emission from the \psr\ system
\citep{mcb+04,pdm+04,cpb04} is particularly exciting.  Energetic
pulsars generate several forms of X-ray emission: quasi-blackbody
emission from the cooling neutron star surface and/or from heated
polar caps; pulsed non-thermal emission from the pulsar magnetosphere;
and at larger distances from the pulsar, synchrotron emission from a
pulsar wind nebula (PWN) powered by the relativistic particle outflow.
All of these processes may be taking place in the \psr\ system
\citep[see, e.g.,][]{kpg06}.  Specifically, the X-rays could be pulsed
magnetospheric or thermal emission from pulsar A \citep[as seen for
several other recycled pulsars; see][]{z06}, could originate in the
colliding winds of A\ and B\ \citep{lyutikov04}, or could be produced
by the shock generated when one or both of the pulsar winds interacts
with the interstellar medium \citep{lyutikov04,gm04}.

The electrodynamics of pulsar winds have been studied in considerable
detail through the extended PWNe typically seen around young and/or
high-velocity pulsars \citep{gs06}. In systems such as the Crab
Nebula, the PWN is an expanding synchrotron bubble centered on the
pulsar. Such nebulae act as calorimeters, revealing the geometry and
energetics of the pressure-confined outflow and its termination shock.
However, the termination shocks seen in such PWNe are typically at
distances $\sim10^6 - 10^9 R_{LC}$ from their pulsars (where the light
cylinder radius of a pulsar rotating at a frequency $f$ is $R_{LC}
\equiv c/2 \pi f$).  In contrast, the two neutron stars in the double
pulsar system are separated by $\lesssim 10^3 R_{LC,A}$ and only $6.6
R_{LC,B}$; a termination shock between them can thus probe the
properties of a pulsar's relativistic wind at smaller separations from
the central engine than ever studied before.  Additionally, detection
of an orbital phase dependence in the X-ray emission might be expected
\citep[e.g.,][]{at93}.  Such variability could constrain the geometry
of the emission site, thus providing new insights into the wind
physics close to the pulsar.

Here we report on {\em Chandra} observations of the double pulsar which
have high enough time resolution to test for pulsations from either pulsar
and for orbital variability, the latter of which might be expected in the
bow shock or colliding winds interpretation.  Forming histograms of count
rates as a function of phase, we test for modulated X-ray emission at
the periods of pulsar A and pulsar B, as well as for orbital modulation.


\section{Observations and Data Analysis}

The \psr\ system was observed with the {\em Chandra X-ray Observatory}
using the High Resolution Camera (HRC-S) in ``timing mode'', which
provides the highest available time resolution, with events corrected
for the instrumental wiring error and time-tagged to 16~$\mu$s
accuracy. The observations spanned 10.5 binary orbits
but were split into two segments for spacecraft operational reasons.
The first segment of 55~ks began on 2006 February 28, while the second
segment began $\sim 67$~ks after the end of the first, and spanned
38~ks.  The pulsar system was unambiguously detected as a point source
in both segments, at a position $07\hh37\mm51\fs22$
$-30\dd39\am40\farcs3$ (J2000), consistent with positions previously
determined at X-ray and radio wavelengths \citep{mcb+04,cgb05,ksm+06}
at the $\sim$0\farcs5 pointing accuracy of Chandra.

X-ray photons were extracted from a 1\arcsec\ radius circle at the
detected position of the \psr\ system, and the times of arrival for
the photons were corrected to the solar system barycenter using the
JPL planetary ephemeris DE405. Of the 411 photons extracted, we
estimate that $\sim16$ counts were contributed by the X-ray
background. Of course, we cannot identify which of the extracted
photons came from the background, and nor can we assign photons to the
individual pulsars. Instead, we use {\sc
Tempo}\footnote{http://www.atnf.csiro.au/research/pulsar/timing/tempo}
and timing solutions from \citet{ksm+06} to calculate the binary
orbital phase and the rotational phases of both pulsars A and B at
which each photon was emitted.  In keeping with our request that the
observation be split (if necessary) into integer orbit blocks, the
orbital phase of the extracted photons ranged from 0.840 to 6.854 in
the first segment, and from 0.497 to 4.531 in the second.  Thus only
$\sim$5\% of the orbit was sampled 11 times, while the rest was
uniformly sampled 10 times.  In the analysis reported below, we have
ignored the minor oversampling, but we have verified that discarding
3 detected photons to force uniformity in orbital coverage does
not affect our results.

\section{Pulsations from PSR~\psr A}

Forming a histogram of count rate as a function of the rotational
phase of A, we detect X-ray pulses from pulsar A, as illustrated in
\Fref{A}.  The uncertainties on each bin (here and elsewhere in this
work) are 1$\sigma$ (68\%) confidence intervals estimated according to
\citet{gehrels86}.  The pulsations are double-peaked and deeply
modulated, with a pulsed fraction $f \equiv ({\rm Max} - {\rm
Min})/({\rm Max} + {\rm Min})$ of $0.74^{+0.18}_{-0.14}$.  To estimate
the significance of the detection, we calculate the Pearson $\chi^2$
statistic for the pulse profile with 16 bins (degrees of freedom
$\nu$=15), and find $\chi^2/\nu = 7.05$, corresponding to a
probability of only $10^{-15}$ ($\sim 8\sigma$) that the profile is
drawn from a uniform distribution.

\begin{figure*}
\epsscale{0.9} 
\plotone{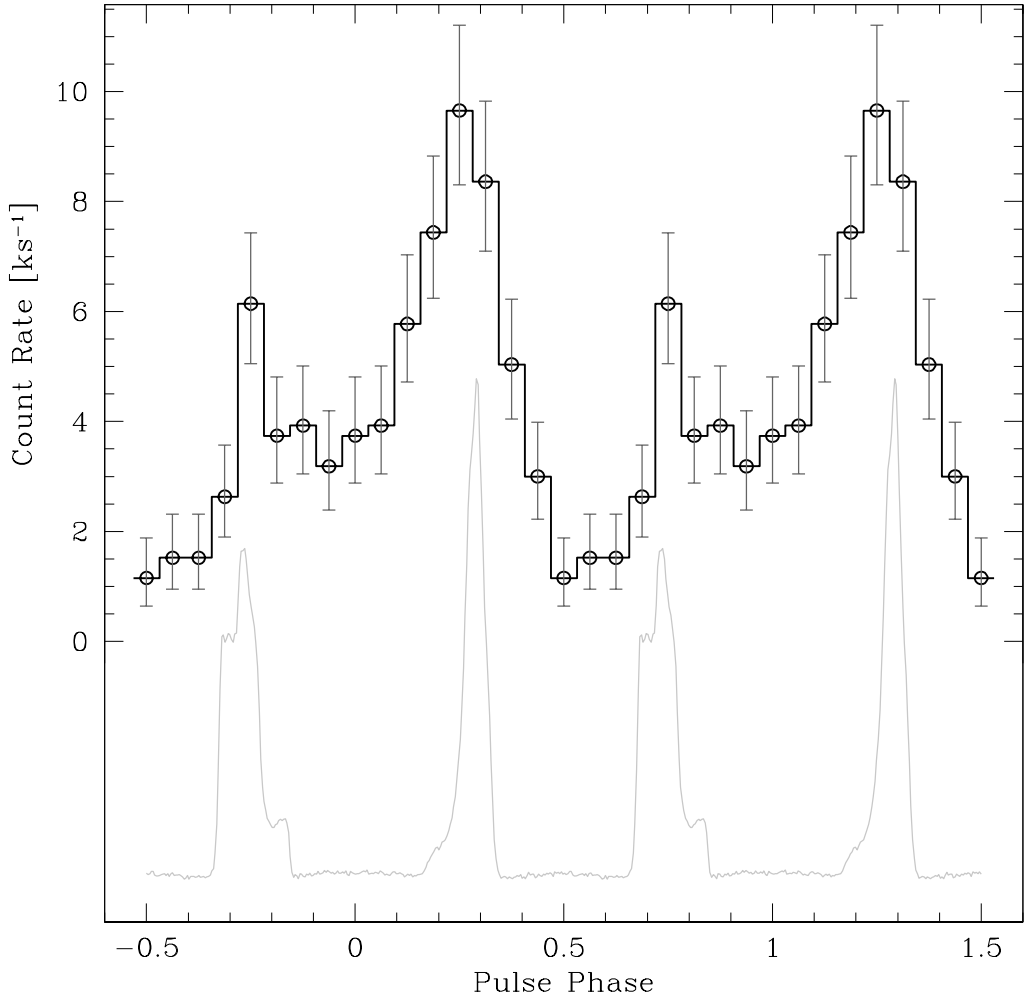}
\caption{X-ray pulse profile of pulsar A, obtained by folding 89~ks of
  {\em Chandra} HRC-S data with {\sc Tempo} and a DE405 timing
  solution.  The uncertainties on each bin (here and elsewhere in this
  work) are 1$\sigma$ (68\%) confidence intervals estimated according
  to \citet{gehrels86}.  The pulse profile is shown twice for clarity,
  and a radio pulse profile obtained at 1.4~GHz \citep{mkp+05} is
  plotted below (in arbitrary units) for comparison.  Both profiles
  are folded using the radio timing solution, and are therefore
  aligned in phase.  }
\label{Fig:A}
\end{figure*} 

A visual comparison of the radio pulse profile of PSR~\psr A
\citep{mkp+05} with the X-ray profile shows a distinct resemblance
(\Fref{A}).
\citet{drb+04} model the radio pulse as two cuts through a wide cone
of emission centered on a single magnetic pole of A, which has its
spin and magnetic axes nearly aligned ($4\arcdeg \pm 3\arcdeg$).
Although a wide range of misalignment is currently permitted by radio
observations \citep{mkp+05}, both peaks of the observed pulse appear
to come from one magnetic pole, implying a very wide fan beam in some
geometries.  The X-ray pulse profile also shows two peaks, whose
locations fall within the range in pulse phase delimited by the radio
peaks when the pulses are phase-aligned.\footnote{\citet{rfk+04} show
that absolute phase alignment is possible at the 60~$\mu$s level
between HRC-S and radio observations of the recycled pulsar
PSR~B1821$-$24. Since we have to predict and account for both orbital
and rotational phase, our timing errors are somewhat larger, but
insignificant compared to the bin width of $\sim$1.4~ms.} This
suggests that the X-ray emission is from a narrower cone than the
radio beam. Specifically, the peaks in the X-ray profile (\Fref{A})
are located at pulse phases $\phi_A \sim 0.27$ and $\sim 0.77$, as
estimated by binning the observed X-ray photons at various
resolutions, and the peak-to-peak separation is $\sim 0.50 \pm 0.01$
($182\arcdeg \pm 3\arcdeg$), while the peaks in the 1.4~GHz radio
profile \citep{mkp+05} are at $\phi_A = 0.234$ and $\phi_A = 0.789$,
separated by $\sim 200$\arcdeg, and the outer rims of the radio
emission profile are at $\phi_A = 0.164$, $0.836$.  The X-ray emission
also shows a significant ``bridge'' between the two peaks, implying
that the cone of X-ray emission is (partially) center-filled in this
model, unlike the broader, hollow radio emission cone.

The detected pulses are quite unlike the typical X-ray emission
observed from other recycled pulsars with comparable spin parameters
\citep[e.g., PSR~J0437$-$4715,][]{zps+02,bgr06,z06}, which show broad,
roughly sinusoidal pulsations with low pulsed fractions and thermal
spectra.  Instead, the pulsations from pulsar A resemble non-thermal
pulses seen only from the most energetic recycled pulsars \citep[e.g.,
PSR~B1821$-$24,][]{rfk+04}, even though A is slower rotating and has a
lower spindown energy loss rate ($\dot{E}_A = 5.9 \times
10^{33}$~erg~\persec). In this context, we note that the formation
scenario for double neutron star binary systems \citep{s04} can result
in a shorter accretion episode and thus a higher surface magnetic
field strength compared to other recycled pulsars.  Pulsar A has an
inferred dipole magnetic field $B_A = 6.4 \times 10^9$~G, comparable to
PSR~B1821$-$24 and the binary pulsar B1534+12, but significantly
higher than other recycled pulsars that show predominantly thermal
X-ray emission.  Both pulsars B1821$-$24 and \psr A also lie above the
death line for curvature radiation estimated by \citet{hum05},
suggesting that the processes that power non-thermal magnetospheric
emission in PSR~B1821$-$24 may also operate for pulsar A, although the
two differ substantially in period and $\dot{E}$.

The absence of any useful energy resolution in {\em Chandra} HRC data
precludes spectral fits to the data, but previous {\em Chandra} ACIS
observations can be well-modeled by a power law with a photon index
$\Gamma \sim 2.9 \pm 0.4$ \citep{mcb+04}, and {\em XMM} data is
well-fit by a power law with a photon index $\Gamma \sim
3.5^{+0.5}_{-0.3}$ \citep{pdm+04}.  Joint fits to the {\em Chandra}
and {\em XMM} data \citep{cpb04} allow for both power law ($\Gamma =
4.2^{+2.1}_{-1.2}$) and thermal black body ($kT_{bb} = 0.20 \pm
0.02$~keV) interpretations. Similar fit parameters ($\Gamma \sim 3$ or
$kT_{bb} \sim 0.2$~keV) were found by \citet{kpg06} as well.
Additionally, \citet{cpb04} show that a two-component fit with a fixed
power law index $\Gamma=2$ and a black body component ($kT_{bb} = 0.16
\pm 0.04$~keV) is consistent with the {\em Chandra} ACIS and {\em XMM}
data, although two components are not statistically required.

The X-ray spectrum, in combination with our detection of sharp,
double-peaked X-ray pulses, is thus consistent with a purely
magnetospheric origin for the X-ray emission, but it is also possible
that the observed X-ray pulsations consist of both non-thermal
magnetospheric emission and broad sinusoidal thermal pulsations from
the hot polar cap.  The pulse profile of A shows a floor of X-ray
emission (\Fref{A}), corresponding to a count rate of $\approx 1.5 \pm
0.6$~cts~ks$^{-1}$ at every phase.  An image of the off-pulse counts
reveals no extended nebular structure, and their distribution is
consistent with the on-pulse photons.  Other recycled pulsars also
show emission at all pulse phases, whether their pulsations are broad
and thermal \citep[e.g., PSR J0437$-$4715,][]{zps+02,bgr06,z06} or
narrower and primarily non-thermal (e.g., PSR~J0218+4232,
\citealt{khv+02}; PSR~B1821$-$24, \citealt{rfk+04}).  Such unpulsed
emission is usually ascribed to thermal X-rays emitted from the
neutron star surface.  Assuming that the entire X-ray flux of the
double pulsar system arises only from the combined thermal and
non-thermal emission from PSR~\psr A, we find that the maximum
amplitude sinusoid $A(1 + \sin 2\pi (\phi-\phi_0))$ that is consistent
with the observed profile at $1\sigma$ could account for as much as
$\sim 60\%$ of the observed X-ray counts, although the actual fraction
is likely to be far lower.  Rotational phase-resolved spectroscopy
with substantially more X-ray counts will be required to verify or
rule out such a two-component model.

In order to investigate possible orbital variations in the X-ray pulse
profile of A, 9-bin pulse profiles were constructed for each quadrant
of the orbit. Each of the four profiles was then compared to the pulse
profile constructed by averaging the other three quadrants. The
resulting $\chi^2/\nu$ values range between 0.8 and 1.3 (with $\nu =
9$ degrees of freedom), consistent with no variations.  While we lack
the S/N to definitively rule out any differences between the X-ray
pulse profiles, no orbital variations are detected in the radio pulse
profiles of A either \citep[e.g.][]{ksm+06}.

We note in passing that our estimate of the pulsed fraction $f =
0.74^{+0.18}_{-0.14}$ is only marginally consistent with the upper
limit of 60\% on the pulsed fraction (assuming sinusoidal pulses)
inferred by \citet{pdm+04} from {\em XMM-Newton} observations.  Since
the detected pulse is non-sinusoidal, a direct comparison is not
possible, but $\sim 60\%$ of our detected photons are above the
estimated minimum count rate baseline, and $\gtrsim 51\%$ are
$>1\;\sigma$ above the baseline level.  The {\em XMM} pn observations
of \citet{pdm+04}, which were in continuous clocking mode, were
totally dominated by the background due to the one-dimensional
readout, while the {\em XMM} MOS chips lack the time resolution to
detect pulses from A, leading to a limit which is less robust compared
to the {\em Chandra} HRC detection presented here.

\section{Non-detection of PSR~\psr B}

We repeated the analysis described in \S~3
for PSR \psr B. The results are shown in \Fref{B}. No X-ray pulsations
are detected, either by folding the full span of data, or by selecting
counts which are in the lowest emission bins of the pulse profile of
A, $0.46875 < \phi_A < 0.65625$.  For the folded profile from the
entire data span, we calculate $\chi^2/\nu = 1.34$, corresponding to a
17\% probability that the data are drawn from a uniform
distribution. As described by \citet{lew83}, epoch folding is not as
sensitive to broad, smooth pulses as the family of Rayleigh statistics
$Z^2_m$, which also avoid the need to bin data.  Therefore, we also
calculated the $H$ statistic \citep[$H \equiv {\rm Max} (Z_m^2 -4m
+4)$, for $1 \le m \le 20$;][]{jsr89}, which is well suited to
searching for an unknown modulation shape.  We find $H = 0.035$, at
$m=1$, corresponding to a null hypothesis probability (i.e., the
probability that we are sampling a uniform distribution) close to
unity.

Pulsar B shows significant enhancements in radio emission at some
parts of its orbit \citep{lbk+04}, but folding X-ray photons selected
from those orbital phase ranges does not show any evidence for
pulsations either.  The non-detection is unsurprising, since pulsar B
has rotational parameters and a spindown energy loss rate ($\dot{E}_B
= 1.7 \times 10^{30}$~erg~\persec) similar to other ``ordinary''
pulsars (ages $\sim 10^6-10^8$~yr), which are not known for their
X-ray emission. The spindown luminosity of pulsar B is only $\sim 3
\times 10^{-4}\,\dot{E}_A$, and so pulsar A is expected to dominate
any X-ray emission from the system.

\begin{figure}
\epsscale{1.0} 
\plotone{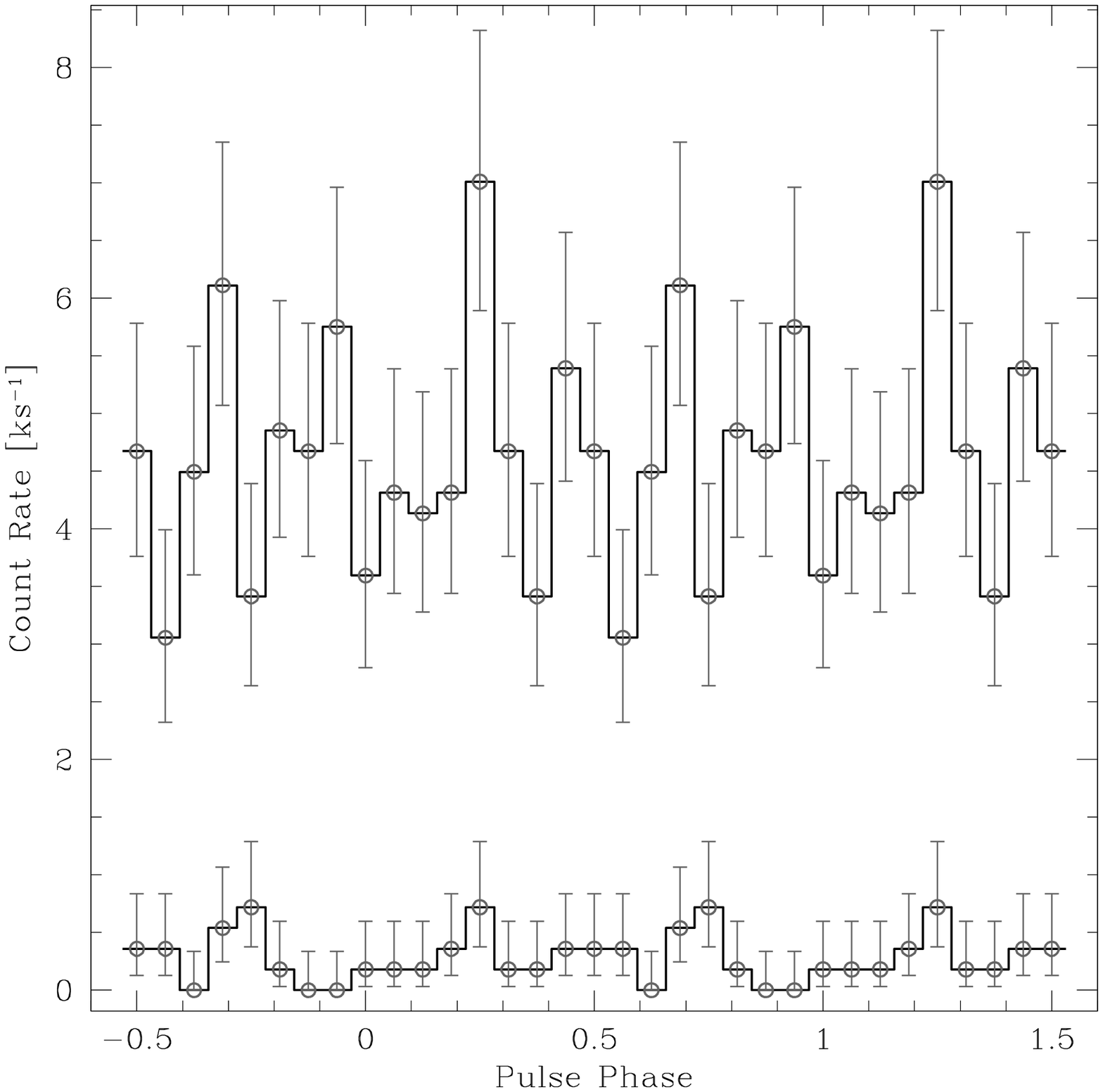}
\caption{Non-detection of pulsations from PSR~\psr B. 89~ks of {\em
  Chandra} HRC-S data were folded at the rotational phase of B, as
  predicted by {\sc Tempo}, but no pulsations were detected. Two pulse
  periods are shown for clarity.  Further, we extracted 26 photons
  detected in the off-pulse phase range of A, $0.46875 < \phi_A <
  0.65625$, corresponding to the three bins with lowest photon counts
  in \Fref{A}. The folded counts are shown on the same scale (the
  lower pulse profile in the figure). Again, no significant pulsations
  were detected.}
\label{Fig:B}
\end{figure} 

\section{Limits on orbital modulation}

Systems such as the Crab nebula \citep{kc84,ga94} and the nebula
around PSR~B1509$-$58 \citep{gak+02} provide the best current
constraints on the behavior of pulsar winds at large distances ($\sim
10^6 - 10^9 R_{LC}$) from the neutron star, but few constraints exist
for the close-in behavior.  The magnetization parameter $\sigma$, the
ratio of Poynting flux to the kinetic energy flux in the wind, is a
key descriptor of such systems.  Optical, near-infrared, and X-ray
images at sub-arcsecond resolution reveal that the shock has an
axisymmetric structure of equatorial arcs (wisps) and polar jets
(knots) that vary on short time-scales \citep{hmb+02,ptks03,msw+05},
and that the wind transforms from a Poynting-dominated outflow
($\sigma \gg 1$) near the pulsar to a kinetic-energy-dominated outflow
($\sigma < 1$) at the termination shock \citep{kc84}.  Recent work has
begun to elucidate the collimation mechanism that produces the
axisymmetric structure \citep[e.g.,][]{kl04}, while the conversion of
Poynting flux to mechanical energy remains poorly understood.

The wind interaction of a neutron star with a stellar binary companion
allows constraints on the wind behavior at $\sim 10^4 R_{LC}$, and
such interaction can produce radio and high energy emission
signatures.  For example, the Be star---pulsar binary B1259$-$63
produces unpulsed radio emission \citep{bmjs99} as well as unpulsed
high energy emission \citep[e.g.,][]{gtp+95}, which arise from the
shock formed between the stellar outflow and the pulsar wind
\citep{ta97}.  The interaction of the pulsar B1957+20 with its white
dwarf binary companion is expected to produce orbital modulation in
the X-ray emission \citep[see, e.g.,][]{at93,michel94}, although the
observational evidence for such modulation \citep{sgk+03,hb07} is not
significant.

As opposed to the interaction between a neutron star relativistic wind
and the particle wind of a stellar companion, the double pulsar
presents a situation where the relativistic wind interacts with the
magnetosphere of another neutron star.  Additionally, the system
separation is $\lesssim 10^3 R_{LC,A}$ and only $6.6 R_{LC,B}$.  The
detection of orbital modulation in the system would thus be of
particular interest, since it probes the behavior of the pulsar wind
in a high-$\sigma$ regime.  We note that \citet{kpg06} find possible
orbital phase dependence for the double neutron star binary B1534+12,
but not for the \psr\ system. Their result is based on earlier {\em
XMM} and {\em Chandra} ACIS data that lacked the time resolution to
detect pulsations from pulsar A.

Given our detection of deeply modulated pulsed emission from PSR~\psr
A, we attempted to detect orbital modulation in the X-ray emission by
folding X-ray photons from the off-pulse phase of A, $0.46875 < \phi_A
< 0.65625$, corresponding to the three bins with lowest photon counts
in \Fref{A}.  The 25 counts thus selected from the observation
(corresponding to a reduced effective exposure of 16.7~ks) were folded
as a function of orbital phase, and the results are shown in the top
panel of \Fref{orbit}.  We find an apparent enhancement at a phase
$\sim$0.69. At the epoch of observation, that phase bin includes 
an orbital longitude $\omega = 0\arcdeg$, corresponding to A's crossing
of the ascending node of the orbit.  However, there is no obvious
physical mechanism that could produce such an enhancement, and the
binned distribution has $\chi^2/\nu = 1.45$, corresponding to a chance
probability of 11.5\%.  As in \S4, we calculate the $H$ statistic with
the un-binned orbital phase values.  We find $H=7.35$ at the fifth
harmonic $Z^2_5$, which allows the null hypothesis that we have
sampled uniformly distributed data at $\sim 5\%$, a probability that
is small but not insignificant.

\begin{figure}
\epsscale{1.2}
\plotone{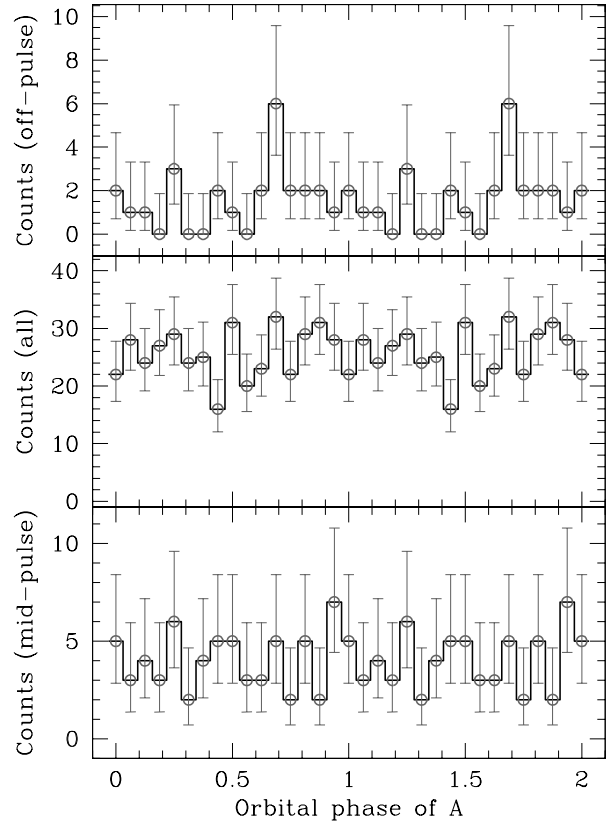}
\caption{ Searching for orbital modulation in X-ray emission from the
  \psr\ system.  In all cases, the orbit is plotted twice for clarity.
  {\em Top:} We extract 25 photons detected in the off-pulse phase
  range of A, $0.46875 < \phi_A < 0.65625$, corresponding to the three
  bins with lowest photon counts in \Fref{A}. Folding at the binary
  phase shows an enhancement at an orbital phase $\sim$0.69.  At the
  epoch of observation, that phase bin encompasses an orbital longitude
  $\omega=0\arcdeg$, when A crosses the ascending node of the orbit.
  {\em Middle:} Folding all detected photons at the binary phase does
  not show such an enhancement.  Note that we lack enough counts to
  detect or constrain the eclipse of A at an orbital longitude $\omega
  = 90\arcdeg$.  {\em Bottom:} We extract 65 photons in the mid-pulse
  of A, $-0.15625 < \phi_A < 0.03125$, corresponding to the three bins
  between the peaks of the profile in \Fref{A}.  The absence of any
  significant enhancement confirms that the apparent signal in the
  top panel is spurious.}
\label{Fig:orbit}
\end{figure}

We also check for the enhancement by folding all the X-ray photons,
and by selecting and folding photons from the bridge of emission
between the two peaks of A's pulse profile ($-0.15625 < \phi_A <
0.03125$), where the contribution of the pulsar itself is reduced. The
results are shown in the middle and bottom panels of \Fref{orbit}, and
in each case, we again calculate the $H$ statistic.  $H = 1.66$ at
$m=1$ when including all the extracted photons, corresponding to a
null hypothesis probability of 52\%.  For the photons chosen from
between the two peaks of A's pulse, $H = 0.023$ at $m=1$, which allows
the null hypothesis at a probability close to unity.  Together, these
results lend weight to the conclusion that the apparent orbital
modulation seen above (with a chance probability of 5\%) is, in fact,
not real.  We have also confirmed the absence of significant
modulation by binning as a function of orbital longitude rather than
phase, with very similar results.  As outlined in \S~3, it is more
likely that the unpulsed X-rays have their origin in thermal emission
from the surface of pulsar A.  We thus confirm the negative result
found by \citet{kpg06}.

From the drifting sub-pulses detected in B's radio emission
\citep{mkl+04}, it is apparent that the low-frequency electromagnetic
wave in the relativistic wind from A influences the emission of pulsar
B, and several models have been proposed where the formation of a
shock between the two pulsars should produce orbital modulation in
their emission \citep[e.g.][]{lyutikov04,gm04,tt04}.

However, only a small fraction of the wind power emitted by A (and
half of the power emitted by B) is intercepted by the shock between
the two pulsars, reducing proportionately the maximum X-ray flux that
the shock emits.  For example, if we assume that the wind energy is
radiated isotropically from A, and that it is intercepted by a sphere
centered on B with radius $R_{LC,B}$, then the power intercepted by
the shock, $\dot{E}_s = 0.006 \dot{E}_A + 0.5 \dot{E}_B \approx
0.006\dot{E}_A$.  If, instead, A's wind is intercepted at the surface
where pressure balance is achieved between the wind from A and the
magnetosphere of B, at $\sim 0.20$ lt-s from B \citep{lbk+04}, then we
have $\dot{E}_s = 0.001 \dot{E}_A + 0.5 \dot{E}_B \approx
0.001\dot{E}_A$.  Finally, if the shock roughly coincides with the
region centered on B that eclipses the radio pulses from A, we have
$\dot{E}_s = 10^{-5} \dot{E}_A + 0.5 \dot{E}_B \approx 1.5 \times
10^{-4} \dot{E}_A$ (although the processes that contribute to radio
eclipses are likely to be quite different from those that cause X-ray
emission).

Of course, the wind radiated from A is unlikely to be isotropic,
especially if the magnetic and rotational axes are nearly-aligned
\citep{drb+04}, and the shock geometry is not described simply by
intersecting spheres centered on A and B.  Nevertheless, the
conservative geometric estimates above demonstrate that the X-ray
power output from the shock is $\dot{E}_s \lesssim 0.006 \dot{E}_A$,
possibly modulated at the orbital period.  Interestingly, spectral
fits to the {\em Chandra} and {\em XMM} data imply an X-ray efficiency
$L_x/\dot{E}_A \lesssim 2 \times 10^{-4}$ \citep{cpb04}, where $L_x$
is the X-ray luminosity in the 0.5---10~keV range.  Thus, if the
entire $\dot{E}_s$ were converted to X-ray emission, at least two 
of our proposed scenarios above would have resulted
in a higher X-ray efficiency for the \psr\ system than actually
observed.  Since we detect X-ray pulses from A which account for a
significant proportion (and arguably $\sim$100\%) of the observed
X-ray emission, all of $\dot{E}_s$ evidently does not appear as X-ray
emission. (We note that for $\dot{E}_A = 5.9 \times
10^{33}$~erg~\persec, the relations derived by \citet{pccm02} for
X-ray luminosity in the 2---10~keV range predict a maximum X-ray
efficiency $L_x/\dot{E}_A < 0.005$, consistent with observations.)

The wind interaction in the double pulsar system is fundamentally
different energetically from wind confinement in a Crab-like pulsar
wind nebula, since the termination shock of the wind is much closer to
pulsar A ($\lesssim 10^3 R_{LC,A}$) than in Crab-like nebulae ($\sim
10^8 R_{LC}$).  All modern wind models, whether for a steady-state,
force-free, magnetohydrodynamic outflow \citep{ck02} or a wave-like,
striped outflow \citep{mm96,lk01}, predict values of the magnetization
parameter $\sigma\gg 1$ (probably $\gtrsim 100$) at these distances,
unlike termination shocks in pulsar wind nebulae, where $\sigma \ll
1$. For a high-$\sigma$ shock, \citet{kc84} estimate an upper limit on
the power fed into the accelerated electrons (and hence on the X-ray
luminosity of the shock) of $\dot{E_s}/(8\sqrt{\sigma})$.  In summary,
as a result of the high expected value of $\sigma$ and the small solid
angle over which the wind from A is intercepted by B, the shock
produced at the interaction region is unlikely to show significant
X-ray emission.  Similar arguments apply to the absence of unpulsed
radio emission from the system \citep{cgb05} as well.

\section{Conclusions}

With 89~ks of {\em Chandra} HRC observations, we have detected deeply
modulated emission from PSR~\psr A.  The off-pulse emission reveals no
extended structure. No pulsations were detected from PSR~\psr B, and
no orbital modulation was detected either.  Although we cannot
absolutely rule out orbital modulation or emission from other
mechanisms such as bow shocks, we have shown that the entire X-ray
emission from the \psr\ system can be explained as arising from pulsar
A alone, either as non-thermal magnetospheric emission, or 
as a combination of magnetospheric and thermal emission.
Pulse phase-resolved spectroscopy will allow discrimination between
these two scenarios.

Like the dog that did not bark in the night, the absence of
significant and detectable orbital modulation in the X-ray emission
from the \psr\ system is noteworthy.  The wind from pulsar A impinges
on and compresses the magnetosphere of B, leading to deep orbital
modulation in the detected radio pulsations from B \citep{lbk+04}, and
the impact of the low-frequency electromagnetic wave in the
relativistic wind from A is also seen in the drifting sub-pulses of
emission observed from B \citep{mkl+04}.  Given the strong influence
of A on the radio emission from B, it may seem natural to ascribe the
X-ray emission from the PSR~\psr\ system to a particle shock formed at
the wind-magnetosphere interaction site.  As we show here, such an
interpretation is neither favored by theory, nor required by the X-ray
observations.  Our observations reveal no significant orbitally
modulated shock emission, consistent with models for relativistic
winds that require a Poynting-dominated wind close to the
pulsar, and with only a small fraction of the energy outflow from A
interacting with the termination shock.



\acknowledgements We thank Michael Kramer for making current timing
solutions available for PSR~\psr A and B, Dick Manchester for
providing us radio pulse profiles, Scott Ransom for creating {\sc
Presto}, and for his guidance in using it, and Zaven Arzoumanian for
helpful discussions about {\sc Tempo}. We also thank George Pavlov,
Slavko Bogdanov, Nanda Rea, and the anonymous referee for their
helpful comments on the manuscript. SC acknowledges support from the
University of Sydney Postdoctoral Fellowship program.  Support for
this work was provided by NASA through Chandra award GO5-6046X to the
Harvard College Observatory.


\begin{thebibliography}{44}
\expandafter\ifx\csname natexlab\endcsname\relax\def\natexlab#1{#1}\fi

\bibitem[{{Arons} \& {Tavani}(1993)}]{at93}
{Arons}, J., \& {Tavani}, M. 1993, \apj, 403, 249

\bibitem[{{Ball} {et~al.}(1999){Ball}, {Melatos}, {Johnston}, \& {Skj{\ae}
  Raasen}}]{bmjs99}
{Ball}, L., {Melatos}, A., {Johnston}, S., \& {Skj{\ae} Raasen}, O. 1999,
  \apjl, 514, L39

\bibitem[{{Bogdanov} {et~al.}(2006){Bogdanov}, {Grindlay}, \&
  {Rybicki}}]{bgr06}
{Bogdanov}, S., {Grindlay}, J.~E., \& {Rybicki}, G.~B. 2006, \apjl, 648, L55

\bibitem[{{Burgay} {et~al.}(2003){Burgay}, {D'Amico}, {Possenti}, {Manchester},
  {Lyne}, {Joshi}, {McLaughlin}, {Kramer}, {Sarkissian}, {Camilo}, {Kalogera},
  {Kim}, \& {Lorimer}}]{bdp+03}
{Burgay}, M., {D'Amico}, N., {Possenti}, A., {Manchester}, R.~N., {Lyne},
  A.~G., {Joshi}, B.~C., {McLaughlin}, M.~A., {Kramer}, M., {Sarkissian},
  J.~M., {Camilo}, F., {Kalogera}, V., {Kim}, C., \& {Lorimer}, D.~R. 2003,
  \nat, 426, 531

\bibitem[{{Campana} {et~al.}(2004){Campana}, {Possenti}, \& {Burgay}}]{cpb04}
{Campana}, S., {Possenti}, A., \& {Burgay}, M. 2004, \apjl, 613, L53

\bibitem[{{Chatterjee} {et~al.}(2005){Chatterjee}, {Goss}, \&
  {Brisken}}]{cgb05}
{Chatterjee}, S., {Goss}, W.~M., \& {Brisken}, W.~F. 2005, \apjl, 634, L101

\bibitem[{{Contopoulos} \& {Kazanas}(2002)}]{ck02}
{Contopoulos}, I., \& {Kazanas}, D. 2002, \apj, 566, 336

\bibitem[{{de Jager} {et~al.}(1989){de Jager}, {Swanepoel}, \&
  {Raubenheimer}}]{jsr89}
{de Jager}, O.~C., {Swanepoel}, J.~W.~H., \& {Raubenheimer}, B.~C. 1989, \aap,
  221, 180

\bibitem[{{Demorest} {et~al.}(2004){Demorest}, {Ramachandran}, {Backer},
  {Ransom}, {Kaspi}, {Arons}, \& {Spitkovsky}}]{drb+04}
{Demorest}, P., {Ramachandran}, R., {Backer}, D.~C., {Ransom}, S.~M., {Kaspi},
  V., {Arons}, J., \& {Spitkovsky}, A. 2004, \apjl, 615, L137

\bibitem[{{Gaensler} {et~al.}(2002){Gaensler}, {Arons}, {Kaspi}, {Pivovaroff},
  {Kawai}, \& {Tamura}}]{gak+02}
{Gaensler}, B.~M., {Arons}, J., {Kaspi}, V.~M., {Pivovaroff}, M.~J., {Kawai},
  N., \& {Tamura}, K. 2002, \apj, 569, 878

\bibitem[{{Gaensler} \& {Slane}(2006)}]{gs06}
{Gaensler}, B.~M., \& {Slane}, P.~O. 2006, \araa, 44, 17

\bibitem[{{Gallant} \& {Arons}(1994)}]{ga94}
{Gallant}, Y.~A., \& {Arons}, J. 1994, \apj, 435, 230

\bibitem[{{Gehrels}(1986)}]{gehrels86}
{Gehrels}, N. 1986, \apj, 303, 336

\bibitem[{{Granot} \& {M{\'e}sz{\'a}ros}(2004)}]{gm04}
{Granot}, J., \& {M{\'e}sz{\'a}ros}, P. 2004, \apjl, 609, L17

\bibitem[{{Grove} {et~al.}(1995){Grove}, {Tavani}, {Purcell}, {Johnson},
  {Kurfess}, {Strickman}, \& {Arons}}]{gtp+95}
{Grove}, J.~E., {Tavani}, M., {Purcell}, W.~R., {Johnson}, W.~N., {Kurfess},
  J.~D., {Strickman}, M.~S., \& {Arons}, J. 1995, \apjl, 447, L113+

\bibitem[{{Harding} {et~al.}(2005){Harding}, {Usov}, \& {Muslimov}}]{hum05}
{Harding}, A.~K., {Usov}, V.~V., \& {Muslimov}, A.~G. 2005, \apj, 622, 531

\bibitem[{{Hester} {et~al.}(2002){Hester}, {Mori}, {Burrows}, {Gallagher},
  {Graham}, {Halverson}, {Kader}, {Michel}, \& {Scowen}}]{hmb+02}
{Hester}, J.~J., {Mori}, K., {Burrows}, D., {Gallagher}, J.~S., {Graham},
  J.~R., {Halverson}, M., {Kader}, A., {Michel}, F.~C., \& {Scowen}, P. 2002,
  \apjl, 577, L49

\bibitem[{{Huang} \& {Becker}(2007)}]{hb07}
{Huang}, H.~H., \& {Becker}, W. 2007, \aap, 463, L5

\bibitem[{{Kargaltsev} {et~al.}(2006){Kargaltsev}, {Pavlov}, \&
  {Garmire}}]{kpg06}
{Kargaltsev}, O., {Pavlov}, G.~G., \& {Garmire}, G.~P. 2006, \apj, 646, 1139

\bibitem[{{Kennel} \& {Coroniti}(1984)}]{kc84}
{Kennel}, C.~F., \& {Coroniti}, F.~V. 1984, \apj, 283, 694

\bibitem[{{Komissarov} \& {Lyubarsky}(2004)}]{kl04}
{Komissarov}, S.~S., \& {Lyubarsky}, Y.~E. 2004, \mnras, 349, 779

\bibitem[{{Kramer} {et~al.}(2006){Kramer}, {Stairs}, {Manchester},
  {McLaughlin}, {Lyne}, {Ferdman}, {Burgay}, {Lorimer}, {Possenti}, {D'Amico},
  {Sarkissian}, {Hobbs}, {Reynolds}, {Freire}, \& {Camilo}}]{ksm+06}
{Kramer}, M., {Stairs}, I.~H., {Manchester}, R.~N., {McLaughlin}, M.~A.,
  {Lyne}, A.~G., {Ferdman}, R.~D., {Burgay}, M., {Lorimer}, D.~R., {Possenti},
  A., {D'Amico}, N., {Sarkissian}, J.~M., {Hobbs}, G.~B., {Reynolds}, J.~E.,
  {Freire}, P.~C.~C., \& {Camilo}, F. 2006, Science, 314, 97

\bibitem[{{Kuiper} {et~al.}(2002){Kuiper}, {Hermsen}, {Verbunt}, {Ord},
  {Stairs}, \& {Lyne}}]{khv+02}
{Kuiper}, L., {Hermsen}, W., {Verbunt}, F., {Ord}, S., {Stairs}, I., \& {Lyne},
  A. 2002, \apj, 577, 917

\bibitem[{{Leahy} {et~al.}(1983){Leahy}, {Elsner}, \& {Weisskopf}}]{lew83}
{Leahy}, D.~A., {Elsner}, R.~F., \& {Weisskopf}, M.~C. 1983, \apj, 272, 256

\bibitem[{{Lyne} {et~al.}(2004){Lyne}, {Burgay}, {Kramer}, {Possenti},
  {Manchester}, {Camilo}, {McLaughlin}, {Lorimer}, {D'Amico}, {Joshi},
  {Reynolds}, \& {Freire}}]{lbk+04}
{Lyne}, A.~G., {Burgay}, M., {Kramer}, M., {Possenti}, A., {Manchester}, R.~N.,
  {Camilo}, F., {McLaughlin}, M.~A., {Lorimer}, D.~R., {D'Amico}, N., {Joshi},
  B.~C., {Reynolds}, J., \& {Freire}, P.~C.~C. 2004, Science, 303, 1153

\bibitem[{{Lyubarsky} \& {Kirk}(2001)}]{lk01}
{Lyubarsky}, Y., \& {Kirk}, J.~G. 2001, \apj, 547, 437

\bibitem[{{Lyutikov}(2004)}]{lyutikov04}
{Lyutikov}, M. 2004, \mnras, 353, 1095

\bibitem[{{Manchester} {et~al.}(2005){Manchester}, {Kramer}, {Possenti},
  {Lyne}, {Burgay}, {Stairs}, {Hotan}, {McLaughlin}, {Lorimer}, {Hobbs},
  {Sarkissian}, {D'Amico}, {Camilo}, {Joshi}, \& {Freire}}]{mkp+05}
{Manchester}, R.~N., {Kramer}, M., {Possenti}, A., {Lyne}, A.~G., {Burgay}, M.,
  {Stairs}, I.~H., {Hotan}, A.~W., {McLaughlin}, M.~A., {Lorimer}, D.~R.,
  {Hobbs}, G.~B., {Sarkissian}, J.~M., {D'Amico}, N., {Camilo}, F., {Joshi},
  B.~C., \& {Freire}, P.~C.~C. 2005, \apjl, 621, L49

\bibitem[{{McLaughlin} {et~al.}(2004{\natexlab{a}}){McLaughlin}, {Camilo},
  {Burgay}, {D'Amico}, {Joshi}, {Kramer}, {Lorimer}, {Lyne}, {Manchester}, \&
  {Possenti}}]{mcb+04}
{McLaughlin}, M.~A., {Camilo}, F., {Burgay}, M., {D'Amico}, N., {Joshi}, B.~C.,
  {Kramer}, M., {Lorimer}, D.~R., {Lyne}, A.~G., {Manchester}, R.~N., \&
  {Possenti}, A. 2004{\natexlab{a}}, \apjl, 605, L41

\bibitem[{{McLaughlin} {et~al.}(2004{\natexlab{b}}){McLaughlin}, {Kramer},
  {Lyne}, {Lorimer}, {Stairs}, {Possenti}, {Manchester}, {Freire}, {Joshi},
  {Burgay}, {Camilo}, \& {D'Amico}}]{mkl+04}
{McLaughlin}, M.~A., {Kramer}, M., {Lyne}, A.~G., {Lorimer}, D.~R., {Stairs},
  I.~H., {Possenti}, A., {Manchester}, R.~N., {Freire}, P.~C.~C., {Joshi},
  B.~C., {Burgay}, M., {Camilo}, F., \& {D'Amico}, N. 2004{\natexlab{b}},
  \apjl, 613, L57

\bibitem[{{McLaughlin} {et~al.}(2004{\natexlab{c}}){McLaughlin}, {Lyne},
  {Lorimer}, {Possenti}, {Manchester}, {Camilo}, {Stairs}, {Kramer}, {Burgay},
  {D'Amico}, {Freire}, {Joshi}, \& {Bhat}}]{mll+04}
{McLaughlin}, M.~A., {Lyne}, A.~G., {Lorimer}, D.~R., {Possenti}, A.,
  {Manchester}, R.~N., {Camilo}, F., {Stairs}, I.~H., {Kramer}, M., {Burgay},
  M., {D'Amico}, N., {Freire}, P.~C.~C., {Joshi}, B.~C., \& {Bhat}, N.~D.~R.
  2004{\natexlab{c}}, \apjl, 616, L131

\bibitem[{{Melatos} \& {Melrose}(1996)}]{mm96}
{Melatos}, A., \& {Melrose}, D.~B. 1996, \mnras, 279, 1168

\bibitem[{{Melatos} {et~al.}(2005){Melatos}, {Scheltus}, {Whiting},
  {Eikenberry}, {Romani}, {Rigaut}, {Spitkovsky}, {Arons}, \& {Payne}}]{msw+05}
{Melatos}, A., {Scheltus}, D., {Whiting}, M.~T., {Eikenberry}, S.~S., {Romani},
  R.~W., {Rigaut}, F., {Spitkovsky}, A., {Arons}, J., \& {Payne}, D.~J.~B.
  2005, \apj, 633, 931

\bibitem[{{Michel}(1994)}]{michel94}
{Michel}, F.~C. 1994, \apj, 431, 397

\bibitem[{{Pavlov} {et~al.}(2003){Pavlov}, {Teter}, {Kargaltsev}, \&
  {Sanwal}}]{ptks03}
{Pavlov}, G.~G., {Teter}, M.~A., {Kargaltsev}, O., \& {Sanwal}, D. 2003, \apj,
  591, 1157

\bibitem[{{Pellizzoni} {et~al.}(2004){Pellizzoni}, {De Luca}, {Mereghetti},
  {Tiengo}, {Mattana}, {Caraveo}, {Tavani}, \& {Bignami}}]{pdm+04}
{Pellizzoni}, A., {De Luca}, A., {Mereghetti}, S., {Tiengo}, A., {Mattana}, F.,
  {Caraveo}, P., {Tavani}, M., \& {Bignami}, G.~F. 2004, \apjl, 612, L49

\bibitem[{{Possenti} {et~al.}(2002){Possenti}, {Cerutti}, {Colpi}, \&
  {Mereghetti}}]{pccm02}
{Possenti}, A., {Cerutti}, R., {Colpi}, M., \& {Mereghetti}, S. 2002, \aap,
  387, 993

\bibitem[{{Rutledge} {et~al.}(2004){Rutledge}, {Fox}, {Kulkarni}, {Jacoby},
  {Cognard}, {Backer}, \& {Murray}}]{rfk+04}
{Rutledge}, R.~E., {Fox}, D.~W., {Kulkarni}, S.~R., {Jacoby}, B.~A., {Cognard},
  I., {Backer}, D.~C., \& {Murray}, S.~S. 2004, \apj, 613, 522

\bibitem[{{Stairs}(2004)}]{s04}
{Stairs}, I.~H. 2004, Science, 304, 547

\bibitem[{{Stappers} {et~al.}(2003){Stappers}, {Gaensler}, {Kaspi}, {van der
  Klis}, \& {Lewin}}]{sgk+03}
{Stappers}, B.~W., {Gaensler}, B.~M., {Kaspi}, V.~M., {van der Klis}, M., \&
  {Lewin}, W.~H.~G. 2003, Science, 299, 1372

\bibitem[{{Tavani} \& {Arons}(1997)}]{ta97}
{Tavani}, M., \& {Arons}, J. 1997, \apj, 477, 439

\bibitem[{{Turolla} \& {Treves}(2004)}]{tt04}
{Turolla}, R., \& {Treves}, A. 2004, \aap, 426, L1

\bibitem[{{Zavlin}(2006)}]{z06}
{Zavlin}, V.~E. 2006, \apj, 638, 951

\bibitem[{{Zavlin} {et~al.}(2002){Zavlin}, {Pavlov}, {Sanwal}, {Manchester},
  {Tr{\"u}mper}, {Halpern}, \& {Becker}}]{zps+02}
{Zavlin}, V.~E., {Pavlov}, G.~G., {Sanwal}, D., {Manchester}, R.~N.,
  {Tr{\"u}mper}, J., {Halpern}, J.~P., \& {Becker}, W. 2002, \apj, 569, 894

\end{thebibliography}
\end{document}